\documentclass[a4paper,10pt,english]{article}
\usepackage[italian]{babel}
\usepackage[latin1]{inputenc}
\usepackage{latexsym}
\usepackage{natbib}
\usepackage{bbm}
\usepackage{amsmath,amssymb,amsfonts}
\usepackage{graphicx}
\usepackage{graphics}
\usepackage{wrapfig}
\usepackage{amsmath}
\usepackage[a4paper]{geometry}
\begin{document}

\title{\center \Large Reflected Backward SDE approach to the price-hedge of defaultable claims with contingent switching CSA }

\maketitle

\begin{center}
  GIOVANNI MOTTOLA\footnote{Mail: g.mottola@be-tse.it}\\
  Sapienza University of Rome
\end{center}
\begin{center}
\textrm{Abstract}\\
\end{center}
\noindent

\noindent In this work we study the \emph{price-hedge} issue for general defaultable contracts characterized by the presence of a \emph{contingent CSA of switching type}.  This is a contingent risk mitigation mechanism that allow the counterparties of a defaultable contract to switch from zero to full/perfect collateralization  and switch back whenever until maturity $T$ paying some \emph{instantaneous switching costs }, taking in account in the picture CVA, collateralization and the funding problem.
We have been lead to the study of this theoretical pricing/hedging problem, by the  economic significance of this type of mechanism which allows a greater flexibility in managing all the defaultable contract risks with respect to the ""standard'' non contingent mitigation mechanisms (as full or partial collateralization).
In particular, our approach  through \emph{hedging strategy decomposition} of the claim (proposition 2.2.5)   and its price-hedge representation through system of nonlinear \emph{reflected BSDE} (theorem 3.2.4) are the main contribution of the work.


\section{\Large Introduction}
\subsection{  Work aim  and contribution}

In this paper we analyze a theoretical contract in which counterparties want to set a contingent CSA (\emph{credit support annex})  in order to gain  the flexibility and the possibility to manage optimally the counterparty risk. 
This is done through a contingent risk mitigation mechanism that allows the counterparties of the contract to switch from zero to full/perfect collateralization and switch back whenever until maturity $T$ paying some \emph{switching costs }and taking into account the \emph{ counterparty risk hedging costs} and - by the other side - the \emph{collateral and funding/liquidity costs} that emerge when collateralization is active. 
Motivated by the economic significance in terms of cost minimization of the studied switching type of mitigation mechanism with respect to the standard - \emph{zero and full collateralization} -  ones, in this work we tackle the \emph{price-hedge}  issue of this generalized defaultable claim with contingent CSA scheme of switching type. \\
This is an hard problem mainly from a theoretical - but also numerical - point of view. The main reasons that make valuation complex are the following.
\begin{description}
  \item[a)] \emph{Recursive valuation problem}: this is the main technical issue to tackle in order to get the price-hedge of the contract. The presence of a CSA in the contract and clearly  of (bilateral) CVA whose value depends on future contract exposures impose to solve the recursion problem through stochastic technique like the \emph{Snell envelope} and the theory of \emph{reflected backward stochastic differential equations } - which we show to be deeply connectes. In particular, the presence of the contingent CSA makes the value of the contract, and in particular of CVA and the collateral/funding term, dependent from the counterparty optimal switching strategy\footnote{See section 3.4 of Mottola (2013)about the issue  of \emph{double order of recursion}.}.
  \item[b)] \emph{Market incompleteness and pricing measure choice}: in pricing models with counterparty risk one typically works under the martingale pricing measure $\mathbb{P}$, but from market incompleteness - due mainly to the unheadgeable \emph{default risk} -  we know that this probability measure is not unique\footnote{From the well known \emph{second fundamental theorem of arbitrage pricing}}. This obviously implies that no perfect hedge can be build and alternative  approaches like mainly the \emph{mean-variance hedging } or \emph{utility/preference based} has to be used.
\end{description}
In the next sections we try to address these issues by setting a fairly general framework (section 2.1) generalizing the BCVA and collateralization to the contingent case and including the funding, assuming the presence of a default free ($\lambda = 0$) external funder in the scheme.
We then study (section 2.2) the   \emph{hedging strategy decomposition} of the whole contract  and we address the recursion issue of the price process by minimizing the relative error of the hedging decomposition from the contract value  in the variance minimization sense. We show (section 3.1) that this issue has a general representation through system of nonlinear \emph{reflected BSDE} which is the main contribution of the work (as resumed in theorem 3.2).

\subsection{ Literature references}

As concerns the main literature references, as regards the framework we refer  to Mottola (2013) and to the reference therein.
In particular are worth of mention the works of Cesari (2010), Gregory (2009), and Brigo, Capponi \emph{et al.} (2011) on the impact of collateralization (including rehypotecation and netting)  on CVA and over range of defaultable claims in the arbitrage-free pricing context, and \emph{Brigo et al.} (2011) that generalizes the CVA valuation in presence of CSA including  funding, the so called funding value adjustment (\emph{FVA}).
As already mentioned in the former section, the main research issue here is to derive the price-hedge of  our whole contract including the contingent CSA. As far as we know, the work of Crepey (2011) has been one of the first rigorous approach to pricing and hedging of general defaultable contract in presence of CSA and funding.\\
The main results of this important work are the derivation, in a reduced form setting under  \emph{pre default} valued( $\mathcal{F}$-adapted or predictable) processes and markovian dynamics for the driving stochastic factors, of a markovian BSDE as the main tool to price and hedge the general CVA\footnote{Actually they derive the BSDE also for the whole contract price, but once one knows the CVA term can derive the other and viceversa. } process of the contract (including CSA and funding).\\
In addition, although the existence of the solution of this (non standard) \emph{backward SDE} (BSDE) is assumed (not proved), thanks to well-known results of \emph{El Karoui et al.} (1997) on BSDE representation the author is able to derive the corresponding \emph{partial-integro differential equation} representation - given the assumption of a markovian jump diffusion for the dynamic - and the explicit form of the hedge process, which is optimal in \emph{min-variance} sense, namely the hedge minimizes the market risk variance due to the hedging error/costs related to CVA. For the issues related to the hedging of defaultable claims we refer also to Bielecki et al. (2004) and for the BSDE theory to Pham (2009) and Yong and  Zhu (1999). \\
In this work we follow in particular Crepey (2011) in order to tackle the price-hedge problem for our generalized theoretical contract, which is  complicated by the  CSA contingency, namely its price  process depends on the optimal switching strategy through time.

\section{\Large BSDE and the price/hedging issue for a general contract with contingent collateralization }

\subsection{Framework and assumptions}

Let us start by stating all the framework characteristics and working assumptions, giving  the definitions  of the processes and variables involved. In the following we keep the perspective of one of the counterparties of our defaultable contract with contingent CSA, given that we are tackling a pricing problem in the arbitrage-free context.
Let's consider a probability space described by the triple $(\Omega, \mathcal{G}_{t}, \mathbb{P})$ where the full filtration is given by  $\mathcal{G}_{t} = \sigma(\mathcal{F}_{t} \vee \mathcal{H}^{A}\vee\mathcal{H}^{B})_{t\geq0}$ for $t\in[0,T]$  and $\mathbb{P}$  the real probability measure defined on this space. On it lie two strictly positive random time $\tau_{i}$ for $i\in \{A,B\}$, which represent the \emph{default times} of the counterparties considered in our model.  In addition, we define the \emph{default process } $H^{i}_{t} = \mathbbm{1}_{\{\tau^{i} \leq t\}}$ and the relative filtration $\mathcal{H}^{i}$ generated by $H_{t}^{i}$ for any $t\in \mathbb{R}^{+}$. We are left to mention $\mathcal{F}$ which is the (risk-free) \emph{market filtration} generated by a $d$-dimensional Brownian motion vector $W$  adapted to it, under the real measure $\mathbb{P}$. In addition, we remember that all the processes we consider, in particular $H^{i}$,  are \emph{c\'adl\'ag semimartingales} $\mathcal{G}$ adapted and $\tau^{i}$ are $\mathcal{G}$ stopping times.\\ 
For convenience, let us define the first default time of the counterparties as $\tau = \tau_{A}\wedge\tau_{B}  $ which also represent the ending/exstinction time of the underlying contract, with the corresponding indicator process $H_{t} = \mathbbm{1}_{\{\tau\leq t\}}$.
As concerns the underlying market model it is assumed arbitrage-free, namely it admits a \emph{spot martingale measure} $\mathbb{Q}$ (not necessarily unique) equivalent to $\mathbb{P}$. A spot martingale measure is associated
with the choice of the savings account $B_{t}$ (so that  $B^{-1}$ as discount factor) as a numeraire that, as usual,  is given by $\mathcal{F}_{t}$-predictable process
\begin{equation}
    B_{t}= \exp \int_{0}^{t}r_{s}ds, \:\:\forall \: t \in \mathbb{R}^{+}
\end{equation}
where the short-term $r$ is assumed to follow an $\mathcal{F}$-progressively measurable stochastic process (whatever it  is the choice of the term structure model for itself).\\
We then define the \emph{Az\'ema supermartingale} $G_{t} = \mathbb{P}(\tau > t |\mathcal{F}_{t})$ with $G_{0}=1$ and $G_{t} >0\:\: \forall \: t \in \mathbb{R}^{+} $  as the \emph{survival process} of the default time $\tau$ with respect to the filtration $ \mathbb{F}$. 
The process $G$ being a bounded $\mathcal{G}$-supermartingale it admits a unique \emph{Doob-Meyer decomposition} $G = \mu - \nu$, where $\mu$ is the martingale part and $  \nu$ is a predictable increasing process. In particular, $\nu$ is assumed to be absolutely continuous  with respect to the \emph{Lebesgue measure}, so  that $d\nu_{t} = \upsilon_{t} dt$ for some $\mathcal{F}$-progressively measurable, non-negative process $\upsilon$.
So that, we can define now the default intensity $\lambda$ as the $\mathcal{F}$-progressively measurable process that is set as $\lambda_{t} = G_{t}^{-1}\upsilon_{t}$ so that $dG_{t} = d\mu_{t} - \lambda_{t} G_{t}dt$ and the cumulative default intensity is defined as follows
\begin{equation}
    \Lambda_{t}=\int_{0}^{t} G_{u}^{-1} d\nu_{u} = \int_{0}^{t} \lambda_{u}du,
\end{equation}
For convenience, we assume that the \emph{immersion property} holds in our framework, so that every c\'adl\'ag $\mathcal{G}$-adapted (square-integrable) martingale is also $\mathcal{F}$-adapted. 


In particular, we assume \emph{pre-default valued processes} namely, setting $J := 1-H = \mathbbm{1}_{\{t\leq \tau \}}$ we have that for any $\mathcal{G}$-adapted, respectively $\mathcal{G}$-predictable process $X$ over $[0,T]$, there exists a unique $\mathcal{F}$-adapted,  respectively $\mathcal{F}$-predictable,  process $\tilde{X}$ over $[0,T]$, called the pre-default value process of $X$, such that $JX = J\tilde{X}$, respectively $J_{-}X = J_{-}\tilde{X}$.\\

Let us now pass to define the main processes and variables involved in the analysis of the price-hedge for our contingent claim. In particular, we recall here the bilateral CVA definition which is generalized to the contingent CSA case. Starting by default-free and defaultable dividend and price processes, we state the following definitions.\\

\textbf{Definition 2.1.2 (Clean dividend and price process).}
\emph{The clean dividend process $D^{rf}_{t}$ of a counterparty default free (exchange traded)  contract is the $\mathcal{F}$-adapted process described by the final payoff $X$, the cashflows $A$ and $\tau =\tau^{i}=\infty$ as follows
\begin{eqnarray}
  D_{t}^{rf} &=& X_{T}  +\sum_{i\in\{A,B\}}\bigg(\int_{]t, T]} d\mathbf{A}^{i}_{u} \bigg)  \;\;t\in[0,T].
\end{eqnarray}
The clean price process $S^{rf}$ would be simply represented by the integral over time of the dividend process under the relative pricing measure, that is
\begin{equation}
   S_{t}^{rf}= B_{t}\mathbb{E}_{\mathbb{Q}} \bigg( \int_{]t,T]} B_{u}^{-1} dD^{rf}_{u} \big| \mathcal{F}_{t} \bigg)\;\;t\in[0,T].
\end{equation}}

\textbf{Definition 2.1.3 (Bilateral risky dividend and price process).} \emph{The dividend process $D_{t}$ of a defaultable claim with bilateral counterparty risk $(X; \mathbf{A};Z; \tau )$, is defined as
the total cash flows of the claim until maturity $T$
that is formally
\begin{eqnarray}
  D_{t} &=& X_{T} \mathbbm{1}_{\{T<\tau\}}  + \sum_{i\in\{A,B\}}\bigg(\int_{]t, T]}  (1-H_{u}^{i})d\mathbf{A}^{i}_{u} +  \int_{]t,T] } Z_{u}dH_{u}^{i}\bigg)  \;\;t\in[0,T]
\end{eqnarray}
for $i \in \{A,B\}$.\\
Similarly, the (ex-dividend) price process $S_{t}$ of a defaultable claim with bilateral counterparty risk maturing in $T$ is defined as the integral of the discounted  dividend flow under  the risk neutral measure $\mathbb{Q}$, namely the $NPV_{t}$ of the claim, that is formally
\begin{equation}
   NPV_{t} = S_{t}= B_{t}\mathbb{E}_{\mathbb{Q}} \bigg( \int_{]t,T]} B_{u}^{-1} dD_{u} \big| \mathcal{G}_{t} \bigg)\;\;t\in[0,T].
\end{equation}}

The following proposition state. For the proof we refer to Mottola (2013).\\

\textbf{Proposition 2.1.4 (Bilateral CVA).} \emph{The bilateral CVA process of a defaultable claim with bilateral counterparty risk $(X; \mathbf{A};Z; \tau )$ maturing in $T$ satisfies the following relation\footnote{The formulation is seen from the point of view of $B$. Being symmetrical between the party, just the signs change.}
\begin{eqnarray}
  BCVA_{t} &=& CVA_{t} - DVA_{t} \nonumber \\
  &=& B_{t}\mathbb{E}_{\mathbb{Q}} \bigg[ \mathbbm{1}_{\{t< \tau=\tau_{B}\leq T\}}B_{\tau}^{-1}(1-R_{c}^{B})(S^{rf}_{\tau})^{-} \bigg|\mathcal{G}_{t} \bigg] + \nonumber \\
  &-& B_{t}\mathbb{E}_{\mathbb{Q}}\bigg[ \mathbbm{1}_{\{t< \tau=\tau_{A}\leq T\}}B_{\tau}^{-1}(1-R_{c}^{A})(S^{rf}_{\tau})^{+}\bigg|\mathcal{G}_{t}  \bigg]
\end{eqnarray}
for every $t \in[0,T]$, where $R_{c}^{i}$ for $i\in\{A,B\}$ is the counterparty recovery rate (process).}\\

Let us now define the collateral process as follows.\\

\textbf{Definition 2.1.5 (Collateral account/process)}. \emph{Let us define for the bilateral CSA of a contract the positive/negative threshold  with $\Lambda_{i}$ for $ i=\{A,B\}$ and the  positive minimum transfer amount with $MTA$. The collateral process $Coll_{t} : [0,T]\rightarrow \mathbb{R}$ is a stochastic  $\mathcal{F}_{t}$-adapted process\footnote{In the literature, depending on the CSA provisions and modeling choice, the process is also considered $\mathcal{F}_{t}$-predictable or in general adapted to $\mathcal{G}_{t}$.} defined as follows
\begin{equation}
    Coll_{t}=\mathbbm{1}_{\{S^{rf}_{t} > \Lambda_{B} + MTA\}}(S^{rf}_{t} - \Lambda_{B}) + \mathbbm{1}_{\{S^{rf}_{t} < \Lambda_{A} - MTA\}}(S^{rf}_{t} - \Lambda_{A}),
\end{equation}
on the time set $ \{t<\tau\}$, and
    \begin{equation}
    Coll_{t}=\mathbbm{1}_{\{S^{rf}_{\tau^{-}} > \Lambda_{B} + MTA\}}(S^{rf}_{\tau^{-}} - \Lambda_{B}) + \mathbbm{1}_{\{S^{rf}_{\tau^{-}} < \Gamma_{A} - MTA\}}(S^{rf}_{\tau^{-}} - \Lambda_{A}), \;
\end{equation}}
 on the set $ \{\tau\leq t <\tau +\Delta t\}$ (with $\Delta t$  indicating a small time after the default time $\tau$).\\\\

In the \emph{perfect collateralization } case one can easily show that  $Coll^{Perf}_{t}$ is always equal to the mark to market, namely to the (default free) price process $S_{t}^{rf}$ of the underlying claim, that is formally (by taking the definition 2.1.4)
\begin{equation}
    Coll_{t}^{Perf}=\mathbbm{1}_{\{S^{rf}_{t} > 0 \}}(S^{rf}_{t} - 0) + \mathbbm{1}_{\{S_{t}^{rf} <0\}}(S^{rf}_{t} - 0) = S^{rf}_{t} \;\; \forall \:  t \in [0,T], \; on\:\{t<\tau\}.
\end{equation}
and
\begin{equation}
    Coll_{t}^{Perf}=  S^{rf}_{\tau^{-}} \;\; \forall \:  t \in [0,T], on \: \{ \tau\leq t <\tau+\delta t\}
\end{equation}

 Now, in order to generalize the analysis to the contingent CSA, we need to introduce the \emph{switching controls} in the framework. These are the controls, namely the implicit options of the contract, that empower  the counterparty $A$  to optimally manage the counterparty risk by calling the collateral. By the characteristic of the contingent CSA, he can ""give impulse'' to the system every times in $[0,T]$ through a sequence of switching times, say $\tau_{j} \in \mathcal{T}$, with $\mathcal{T}\subset [0,T]$, where we denote the switching set
\begin{equation}
    \mathcal{T}:= \big\{ \tau_{1},\dots, \tau_{M} \} = \big\{ \tau_{j} \big\}_{j=1}^{M}
\end{equation}
with the last switching $\{\tau_{M} \leq T\}$ ($M< \infty$) and the $\tau_{j}$ are, by definition of \emph{stopping times}, $\mathcal{F}_{t}$-measurable random variable. Therefore, at any of these switching times, we need to define the related set of the \emph{switching/impulse indicator}, which is quite trivial in our case with only two possible switching regime:
\begin{equation}
    \mathcal{Z}:= \big\{ z_{1},\dots, z_{M}\} = \big\{ z_{j}\big\}_{j=1}^{M}
\end{equation}
with the  $z_{j}$'s that are  $\mathcal{F}_{\tau_{j}}$-measurable switching indicators taking values $z_{j}=\{ 0,1\} \: \forall j=1,\dots,M $, so we have
\[
 z_{j} =
  \begin{cases}
  1 & \; \Rightarrow\; ''zero\; collateral\;'' \;(full\; CVA)\\
  0      &  \; \Rightarrow\; ''full\; collateral\;'' \;(null\; CVA)
  \end{cases}
\]

So, we have that the control set for our problem is formed of all the sequences of indicators and switching times $ \mathcal{C} \in \mathbb{R}_{M}^{+}$, that is
\begin{equation*}
    \mathcal{C}=\big\{ \mathcal{T}, \mathcal{Z} \big\} = \big\{ \tau_{j}, z_{j}\big\}_{j=1}^{M}
\end{equation*}.\\

We can now give the definitions for contingent collateral and contingent BCVA (referring for the details to section two of Mottola (2013)).\\

 \textbf{Definition 2.1.6 (Contingent Collateral process)}.\emph{The contingent collateral  $Coll^{C}_{t}$ is the $\mathcal{F}_{t}$-adapted process defined for any time $t\in[0,T]$ and  for every switching time $\tau_{j} \in [0,T] $ and $j=1,\dots,M$, switching indicator $z_{j}$ and default time $\tau$ (defined above as $\min\{\tau_{A},\tau_{B}\} $ ), as follows
\[
 Coll^{C}_{t} =
  \begin{cases}
  S^{rf}_{t}\mathbbm{1}_{\{ z_{j}=0 \}} \mathbbm{1}_{\{ \tau_{j} \leq t < \tau_{j+1} \}} + 0 \mathbbm{1}_{\{ z_{j}=1 \}} \mathbbm{1}_{\{ \tau_{j} \leq t < \tau_{j+1} \}}  & \;\: on \: \{ t<\tau\}\\
  S^{rf}_{\tau^{-}}\mathbbm{1}_{\{ z_{j}=0 \}} \mathbbm{1}_{\{ \tau_{j} \leq t < \tau_{j+1} \}} + 0 \mathbbm{1}_{\{ z_{j}=1 \}} \mathbbm{1}_{\{ \tau_{j} \leq t < \tau_{j+1} \}}     &  \;\: on \{ \tau\leq t <\tau+\Delta t\}
  \end{cases}
\]
 where we recall that $S_{t} = Coll_{t}^{Perf} $ from point b) results.}\\

\textbf{Definition 2.1.7 (BCVA with contingent CSA)}. \emph{The bilateral CVA of a contract with contingent CSA of switching type, $BCVA^{C}_{t}$ is the $\mathcal{G}_{t}$-adapted process defined for any time $t\in[0,T]$,  for every switching time $\tau_{j} \in [0,T] $ and $j=1,\dots,M$, switching indicator $z_{j}$ and default time $\tau$ (defined as above), as follows
\begin{equation}
    BCVA^{C}_{t} = BCVA_{t}\mathbbm{1}_{\{ z_{j}=1 \}} \mathbbm{1}_{\{ \tau_{j} \leq t < \tau_{j+1} \}} + 0 \mathbbm{1}_{\{ z_{j}=0 \}} \mathbbm{1}_{\{ \tau_{j} \leq t < \tau_{j+1} \}}, \;\: for \: t\in[ 0,T\wedge\tau],
\end{equation}
where the expression for $BCVA$ is known from Proposition 2.2.3.}\\

From the above definitions of\emph{ contingent CSA} and (bilateral) \emph{CVA } , let us highlight that the switching controls enter and affect the dynamic of these processes . In particular,  switching to full collateralization implies  $BCVA=0$, that is $S_{t}=  S_{t}^{rf} = Coll^{perf}_{t}$. So, formally, we have:\\

$if \: \big\{z_{j}= 1\big \} \: and \: \{\tau_{j}\leq t<\tau_{j+1}\}\Rightarrow \;$
\begin{eqnarray*}
  D_{t}^{C} &=& D_{t} \Rightarrow\\
  S_{t}^{C} &=& S_{t} \Rightarrow\\
  BCVA_{t}^{C} &=& BCVA_{t} \forall\:t\in[0,T\wedge \tau]
\end{eqnarray*}


$if \: \big\{z_{j}= 0 \big \} \; and \; \{\tau_{j}\leq t<\tau_{j+1}\}\Rightarrow $
\begin{eqnarray*}
  D_{t}^{C} &=& D_{t}^{rf} \Rightarrow\\
  S_{t}^{C} &=& S_{t}^{rf} =Coll_{t}^{Perf}  \Rightarrow\\
  BCVA_{t}^{C} &=& 0 \forall\:t\in[0,T\wedge \tau]\\
\end{eqnarray*}


We are left to introduce the funding issue in the picture. As concerns our problem, we assume the existence
of a cash account, say $C^{Fund}_{t}$, hold by an external funder assumed \emph{default-free}. This account can be positive or negative depending on the funding $(> 0)$ or investing $(< 0)$ strategy that counterparty uses in relation
to the price and hedge of claim which is also collateralysed in our case.
In order to reduce the problem of recursion in  relation to CVA, we make the assumption that the CVA value is charged to the counterparty but no hedging portfolio is set, so that the funding costs are null $C^{Fund} = 0$ (namely the funding account is not active)\footnote{An alternative simplifying assumption is that the CVA hedging is realized without resorting to funding. }.\\
As regards the funding issue in  relation to the collateralization, under the assumptions of segregation (\emph{no rehypo}) and collateral made up by cash, we distinguish the following cases:
\begin{description}
  \item[1)] If the counterparty has to post collateral in the margin account, she sustains a funding cost, applied by the external funder, represented by the \emph{borrowing rate} $r^{borr}_{t} = r_{t} + s_{t}$ that is the risk free rate plus a credit spread (s) (that is usually different from the other party) . By the other side, the counterparty receives by the funder  the remuneration on the collateral post, that is usually defined in the CSA as a risk free rate plus some basis points, so that we can approximate it at the risk free rate, that is $r_{t} + bp_{t} \cong r_{t}$.\\
      Clearly, we remark that the existence of the above defined rate implies the existence of the following funding assets
      \begin{eqnarray}
         dB^{borr}_{t} &=& (r_{t} + s_{t})B^{borr}_{t}dt  \\
        dB^{rem}_{t} &=& (r_{t}+bp_{t} )B^{rem}_{t}dt
      \end{eqnarray}
  \item[2)] Instead, let's consider the counterparty that call the collateral: as above the collateral is remunerated at the rate given (as the remuneration for the two parties can be different) by $\bar{B}^{rem}$, but she cannot use or invest the collateral amount (that is segregated), so she  sustains an \emph{opportunity cost}, that can be represented by the rate $r^{opp}_{t} = r_{t} + \pi_{t}$, where $\pi$ is a premium over the risk free rate. \\
      Hence, we assume the existence of the following assets too
      \begin{eqnarray}
         dB^{opp}_{t} &=& (r_{t} + \pi_{t})B^{opp}_{t}dt  \\
        d\bar{B}^{rem}_{t} &=& (r_{t}+\bar{bp}_{t} )\bar{B}^{rem}_{t}dt
      \end{eqnarray}
\end{description}

Let us end this section by  resuming our model working assumptions as follows:
\begin{description}
   \item[\textbf{Hp 1)}] \emph{The analysis is focussed on the full/perfect collateralization case (which is set unilaterally, hold by counterparty $A$);}
  \item[\textbf{Hp 2)}] \emph{The CVA cost process is not funded,   $C^{Fund} = 0$;}
  \item[\textbf{Hp 3)}] \emph{All the processes considered are intended to be pre-default value c\'adl\'ag processes (as from lemma 2.1.1.)}
\end{description}

\subsection{Wealth process and hedging strategy decomposition}

Let us  recall  from \emph{definitions 2.1.6} and \emph{2.1.7} on \emph{contingent CSA} that, depending on switching regime $z_{j}$, we have or zero collateral and full bilateral CVA or perfect collateralization (and a null CVA term), namely
\begin{eqnarray*}
  BCVA_{t}^{C}  &+& Coll_{t}^{C} := \\
 BCVA_{t}\mathbbm{1}{\{z_{j}=1 \}}\mathbbm{1}{\{\tau_{j}\leq t< \tau_{j+1} \}}  &+& S_{t}^{rf}\mathbbm{1}{\{z_{j}=0 \}}\mathbbm{1}{\{\tau_{j}\leq t< \tau_{j+1} \}}
\end{eqnarray*}
for all $\tau_{j}, z_{j} \in u$ and $\tau_{j}\in[0,T\wedge \tau]$. This implies, from the contingent price process $S_{t}^{C}$ and $BCVA_{t}^{C}$ relations that

\begin{eqnarray}
S_{t}^{C}&=& S_{t} := S_{t}^{rf} + BCVA_{t} \qquad if\; \{z_{j}=1\}\; and\; \{\tau_{j}\leq t< \tau_{j+1} \}\\
S_{t}^{C}&=&S_{t} := S_{t}^{rf} \qquad \qquad \;\;\;\;\;\;\;\;  if\; \{z_{j}=0\}\; and\; \{\tau_{j}\leq t< \tau_{j+1} \}.
\end{eqnarray}

Actually, by the presence of a funder, the price process of our contract depends also on current and future funding costs/revenues when collateralization is active. Hence, given the presence funding account $C^{Fund}_{t}$ and recalling - by equations (14)-(17) - that the funding assets are deterministic (for convenience) but the funding costs depends every time (recursively) on the price  process  which is the collateral when $z_{j}=0$, under the assumption of CVA \emph{not funded},  we can restate the latter relation as follows

\[
 S_{t}^{C} =
  \begin{cases}
  S_{t}^{rf} + BCVA_{t} & if\; \{z_{j}=1\}\; and\; \{\tau_{j}\leq t< \tau_{j+1} \}\\
  S_{t}^{rf} + C^{Fund}(t,S_{t}^{rf})      & if\; \{z_{j}=0\}\; and\; \{\tau_{j}\leq t< \tau_{j+1} \}
  \end{cases}
\]

This suggests us that the pricing and hedging problem needs an opportune decomposition in order to find the price-hedge related to the switching regime  defined optimally by the counterparty whose objective becomes now - in this pricing context - to minimize the \emph{hedging error/cost} - its mean or variance - from the value of the whole contract, given the impossibility to  perfectly hedge all the risks  deriving by this general defaultable contract.\\

\textbf{Remarks 2.2.1.} The main theoretical issue here is the existence of a martingale pricing measure $\mathbb{Q}$ such that, for the assumed decomposition of the price process depending on the possible switching regimes, it ensures that the price of the contract remain \emph{arbitrage-free}. As already mentioned, this measure is not unique given the impossibility to build a perfect replicating strategy, but a minimization of the hedging error mean or its variance is the strategy typically followed, with the pricing measure  existence (not unique) that is assumed as given. We also make the same assumption here and hence the basic idea is to solve the pricing problem recursively in order to define the optimal switching strategy which has to \emph{minimize the variance error/costs} of the hedging strategy of the whole contract with contingent CSA.\\

More specifically, in order to determine the optimal hedging strategy of our contract  value, we need to assume - as in Crepey (2011) - the existence of a family of \emph{primary market assets} $\mathcal{A}^{asset}$ to be used as \emph{hedging assets} in addition to the \emph{funding assets} as defined in  section 2.1.
In this family is central the assumption on the presence of CDS freshly emitted or par CDS necessary to hedge the CVA term and the counterparty default risk through a rolling strategy.
In fact, these assets are then used to build a \emph{self-financing hedge portfolio}, say $\mathcal{W}$, in which enter all the returns/costs and dividend gain/loss  deriving from the assets processes  used to  hedge (minimizing) all the risks deriving from our whole contract, that can be resumed as follows:
\begin{itemize}
  \item \emph{market risk}, say $\mathcal{R}^{mkt}$, which can be perfectly hedged through an opportune self-financing strategy involving primary market assets, say $\mathcal{A}^{\mathcal{R}^{mkt}}$;
  \item \emph{counterparty/default risk}\footnote{Here is meant to be included also the unheadgeable \emph{jump risk}.}, say $\mathcal{R}^{def}$, which can only be imperfectly hedged via  self-financing strategy involving primary market assets, denoted as  $\mathcal{A}^{\mathcal{R}^{def}}$;
  \item\emph{ funding risk},say $\mathcal{R}^{fund}$, which - being recursively related to price process  - is also not perfectly hedgeable. Let us denote  $\mathcal{A}^{\mathcal{R}^{fund}}$ with the related hedging asset.
\end{itemize}

\textbf{Remarks 2.2.2.} Let us remark that the above denoted sets of risk sources and hedging assets  are in general not \emph{separated} or \emph{orthogonal} that is there can be some primary market assets that can enter, say,  both the hedging market assets set  $\mathcal{A}^{\mathcal{R}^{mkt}}$ and the hedging funding assets set $\mathcal{A}^{\mathcal{R}^{fund}}$. In addition the risk sources are in general correlated. Hence, formally we can define the market hedging assets set $\mathcal{A}^{asset}$ as
\begin{equation*}
  \mathcal{A}^{asset} :=  \bigcup^{\mathcal{R} } \mathcal{A}^{\mathcal{R}} \supset  \bigcup_{ i}\Big(\mathcal{A}^{\mathcal{R}^{i}}, \; i \in \{mkt, def, fund\} \Big) \\
  \end{equation*}
where we denote here with $\mathcal{R}$ the set of all the risk sources (having no zero intersection) containing our contract risks.
Let us also remark that  greater it is the degree of dependence/correlation between the different risk sources underlying the contract price process, greater it will be also the complexity of the contract hedging strategy.\\

In the following of the section, given the decomposition for $S^{C}_{t}$  derived above, we assume  for a matter of convenience and representation, that:
\begin{description}
  \item[[HP 1)]] \emph{The hedging market assets set  $\mathcal{A}^{\mathcal{R}^{def}}$ and $\mathcal{A}^{\mathcal{R}^{fund}}$ are separated, namely}:
      \begin{equation}
        \mathcal{A}^{\mathcal{R}^{def}} \bigcap \mathcal{A}^{\mathcal{R}^{fund}}  =\emptyset
      \end{equation}
\end{description}
Let us mention that that this assumption is not in contrast with that imposed in section 2.1 on the funding of the CVA, as we will show later.
We could have imposed also $\mathcal{A}^{\mathcal{R}^{mkt}} \bigcap \mathcal{A}^{\mathcal{R}^{def}}  =\emptyset$, given that the assets used to hedge counterparty risk (usually the CDS) are different from those used to hedge market risk. But this would be too restrictive. Basically we allow the use of the same asset to hedge different contract risks. Similar reasoning can be carried on also for $\mathcal{A}^{\mathcal{R}^{mkt}} $ and $\mathcal{A}^{\mathcal{R}^{fund}}$. \\

Now, in  order to define the portfolio/wealth process $\mathcal{W}$  and the related hedge for our contract with contingent switching CSA, the main idea is that, depending on the active switching regime of the CSA,  a different hedging strategy will be set up or down in order to minimize all the risks - $\mathcal{R}^{mkt}$, $\mathcal{R}^{fund}$ and $\mathcal{R}^{def}$ -  that will be indeed decomposed in some way between the two switching regimes.
Without specifying formally - in relation to a given claim - all the hedging assets  dynamics, let us explicit the ingredients of the wealth process recalling that all the assets processes will remain alive until one of the counterparties defaults, namely until $\{\tau \wedge T\}$, where $\tau := \tau^{A} \wedge \tau^{B}$, which implies that the following relations hold for the hedging assets vector
\begin{eqnarray}
  \mathcal{A}_{t}^{\mathcal{R}^{mkt}} &:=& \Big(\mathbbm{1}_{\{t<\tau\}} \mathcal{A}_{t}^{\mathcal{R}^{mkt},i}\Big)_{t>t_{0}}   \;, \forall  i  \in \{1,\dots,l\} \\
  \mathcal{A}_{t}^{\mathcal{R}^{def}} &:=& \Big(\mathbbm{1}_{\{t<\tau\}} \mathcal{A}_{t}^{\mathcal{R}^{def},i}\Big)_{t>t_{0}}   \;, \forall  i  \in \{1,\dots,h\} \\
  \mathcal{A}_{t}^{\mathcal{R}^{fund}}  &:=& \Big(\mathbbm{1}_{\{t<\tau\}} d\mathcal{A}_{t}^{\mathcal{R}^{fund},i}\Big   )_{t>t_{0}}   \;, \forall  i  \in \{1,\dots,s\}
\end{eqnarray}
Each of these asset vectors will contain dividend paying assets for which the following also states:
\begin{equation}
    \mathcal{D}_{t}^{\mathcal{R}^{j}} := \Big(\mathbbm{1}_{\{t<\tau\}} \mathcal{D}_{t}^{\mathcal{R}^{j},i}\Big)_{t>t_{0}}
\end{equation}
for $j$ depending on the specific risk source and $i$ on the number of dividend paying asset used for building the hedge.\\
Recalling also the definitions of saving account $B_{t}$ defined in (1) with risk-free rate $r_{t}$, of funding and collateral assets (14-17), we have all the elements to define as follows the gain processes that enter in the self-financing strategy set up to hedge the contract price process.\\

\textbf{Definition 2.2.3 (Free-risk and risky gain processes)}. \emph{Let us denote with $\Gamma_{t}$ the gain process deriving from all the hedging instrument traded to build $\mathcal{W}_{t}$. Given  $g_{t}$  as the spread over the risk-free rate for the risky asset investments, we define the gain processes related to each hedging asset sets as follows
\begin{eqnarray}
d\Gamma_{t}^{\mathcal{R}^{mkt},i} &=& d\mathcal{A}_{t}^{\mathcal{R}^{mkt},i} - (r_{t}\mathcal{A}_{t}^{\mathcal{R}^{mkt},i})dt + d\mathcal{D}_{t}^{\mathcal{R}^{mkt},i} \;,  \qquad \forall  i  \in \{1,\dots,l\}\\
 d\Gamma_{t}^{\mathcal{R}^{def},i} &=& d\mathcal{A}_{t}^{\mathcal{R}^{def},i} - ((r_{t}+ g_{t}^{i})    \mathcal{A}_{t}^{\mathcal{R}^{def},i})dt + d\mathcal{D}_{t}^{\mathcal{R}^{def},i} \;, \forall  i  \in \{1,\dots,h\}
\end{eqnarray}
where $\Gamma_{0} = 0$ and $t\in[0, \tau \wedge T]$.}\\

Given that we work here under the pricing measure $\mathbb{Q}$, the underlying assumptions regarding these gain processes is the following.
\begin{description}
  \item[[HP 2)]] \emph{The gain processes $\Gamma_{t}^{\mathcal{R}^{mkt},i}$, $\Gamma_{t}^{\mathcal{R}^{def},i}$ of definition 2.2.3 are respectively $\mathbb{R}$-valued $(\mathcal{F},\mathbb{Q})$-martingale and  $(\mathcal{G},\mathbb{Q})$-martingale.}
\end{description}

\textbf{Remarks 2.2.4.} Let us remark  here that funding assets normally do not pay any dividend over time and in our model are taken as deterministic. The big deal is that funding costs recursively affect contract price process which makes the relative hedge dependent by the contract price process over time. This point will be clear and formalized in  the following proposition 2.2.5.\\ Therefore, we remark  that - by switching control set definition (11)-(12) - we do not exclude that the same hedging assets be used to hedge the claim over the different switching  regimes. This implies that we can have, as for the market hedging asset sets (and similarly for the other hedging asset sets),
\begin{equation*}
  \bigg\{\mathcal{A}^{z,\mathcal{R}^{mkt}}_{t} \subseteq \mathcal{A}^{\zeta,\mathcal{R}^{mkt}}_{t}\bigg\}
\end{equation*}
where $\{z,\zeta\} $ are the switching regimes' indicators.
\\

So, we can now formalize in the following proposition the \emph{hedging decomposition of the wealth and price process } depending on the switching regimes chosen optimally by the counterparty of our defaultable contract with contingent switching CSA.\\

\textbf{Proposistion 2.2.5 (Hedging strategy decomposition)}. \emph{Assume that CVA is not  funded and the hypothesis HP 1) and HP 2) as true.  Let us set with
\begin{eqnarray}
\phi^{Z,mkt}_{t} &:=& (\phi^{i,z,mkt}_ {t})_{t>t_{0}}   \;, \forall  i  \in \{1,\dots,l\},\; Z\in\{z,\zeta\}, \\
\phi^{z,def}_{t} &:=& (\phi^{i,z,def}_ {t})_{t>t_{0}}   \;, \forall  i  \in \{1,\dots,h\},
\end{eqnarray}
the real valued $\mathcal{F}_{t}$-predictable hedge vector processes and with $(\epsilon^{z}_{t}:=\epsilon_{t}\mathbbm{1}_{\{z_{j}=1\}},\epsilon^{\zeta}_{t}:=\epsilon_{t}\mathbbm{1}_{\{z_{j}=0\}})$ the real valued $\mathcal{F}_{t}$-adapted hedge error/cost processes, both assumed also square integrable. By control set definition for our contingent CSA of section 2.1 and gain processes definition 2.2.3, the wealth  produced by the  hedge in the two switching regimes admits the following decomposition
\begin{equation}
 d\mathcal{W}^{z}_{t} = - dD^{z}_{t} + r_{t}\mathcal{W}_{t}^{z}dt + \phi^{z,mkt}_{t} d\Gamma_{t}^{\mathcal{R}^{mkt}} + \phi^{z,def}_{t} d\Gamma_{t}^{\mathcal{R}^{def}}
\end{equation}
 if $\;\{z_{j}=1\}$  and $ \{\tau_{j}\leq t< \tau_{j+1} \}$ and
\begin{equation}
 d\mathcal{W}^{\zeta}_{t} = - dD^{\zeta}_{t} + (r_{t}+ f(\mathcal{A}^{\mathcal{R}^{fund}}_{t},\mathcal{W}^{\zeta}_{t},\phi^{\zeta}_{t})) \mathcal{W}^{\zeta}_{t}dt + \phi^{\zeta,mkt}_{t} d\Gamma_{t}^{\mathcal{R}^{mkt}}
\end{equation}
if $\{z_{j}=0\}$  and $ \{\tau_{j}\leq t< \tau_{j+1} \}$, $\forall \tau_{j}, z_{j}\in\{ \mathcal{T},\mathcal{Z}\}$,($j=1,\dots,M$), where we denote with $dD_{t}$ the (pre-default) contract dividend  process and with $f(.)$ the  funding costs function  recursively depending on wealth and the hedge.\\
This becomes for the price process, taking in account the hedge error/cost process $(\epsilon^{z}_{t},\epsilon^{\zeta}_{t})$,
\[
  dS^{C}_{t} =
  \begin{cases}
  - dD^{z}_{t} + r_{t}\mathcal{W}^{z}_{t}dt + \phi^{z,mkt}_{t} d\Gamma_{t}^{\mathcal{R}^{mkt}} + \phi^{z,def}_{t} d\Gamma_{t}^{\mathcal{R}^{def}} +d\epsilon^{z}_{t}  & if\; \{z_{j}=1\} \\
  - dD^{\zeta}_{t} + (r_{t}+ f(\mathcal{A}^{\mathcal{R}^{fund}}_{t},\mathcal{W}^{\zeta}_{t},\phi^{\zeta}_{t}))\mathcal{W}^{\zeta}_{t}dt + \phi^{\zeta,mkt}_{t} d\Gamma_{t}^{\mathcal{R}^{mkt}} + d\epsilon^{\zeta}_{t}   & if\; \{z_{j}=0\}
  \end{cases}
\]
with  $\{\tau_{j}\leq t< \tau_{j+1} \}$ and  - by self-financing condition of the hedging strategy - $\mathcal{W}_{0}=S^{C}_{0}$ (the contract fair value) and $\epsilon^{z}_{0} = \epsilon^{\zeta}_{0} =0$.} \\

\emph{Proof}. 1) Let us prove first that the condition expressed in (20) (HP 1) and on the funding of CVA are necessary. The hedge representation (29)-(30)  is just funded on the modeling assumption of BCVA process not funded. In fact, if the BCVA was funded, HP 1) would not be true in general given that - by funding recursion - the hedge will be made up by both funding risk and default risk hedging assets, which implies a different representation of the wealth/hedge decomposition (29)-(30). Formally, an extra term due to funding will enter equation (29) when $\{z_{j}=1\}$, so that we get
\begin{equation*}
 d\mathcal{W}^{z}_{t} = - dD^{z}_{t} + (r_{t}+ f(\mathcal{A}^{\mathcal{R}^{fund}}_{t},W^{z}_{t},\phi^{z}_{t}))\mathcal{W}_{t}^{z}dt + \phi^{z,mkt}_{t} d\Gamma_{t}^{\mathcal{R}^{mkt}} + \phi^{z,def}_{t} d\Gamma_{t}^{\mathcal{R}^{def}}
\end{equation*}
implying the following generalized representation
\[
  dS^{C}_{t} =
  \begin{cases}
  - dD^{z}_{t} + (r_{t}+ f(\mathcal{A}^{\mathcal{R}^{fund}}_{t},\mathcal{W}^{z}_{t},\phi^{z}_{t}))\mathcal{W}_{t}^{z}dt + \phi^{z,mkt}_{t} d\Gamma_{t}^{\mathcal{R}^{mkt}} + \phi^{z,def}_{t} d\Gamma_{t}^{\mathcal{R}^{def}} +d\epsilon^{z}_{t}  & if\; \{z_{j}=1\} \\
  - dD^{\zeta}_{t} + (r_{t}+ f(\mathcal{A}^{\mathcal{R}^{fund}}_{t},\mathcal{W}^{\zeta}_{t},\phi^{\zeta}_{t}))\mathcal{W}^{\zeta}_{t}dt + \phi^{\zeta,mkt}_{t} d\Gamma_{t}^{\mathcal{R}^{mkt}} + d\epsilon^{\zeta}_{t}   & if\; \{z_{j}=0\}.
  \end{cases}
\]
Clearly being $f(\mathcal{A}^{\mathcal{R}^{def}})(.)$   dependent  also  on the $BCVA$ process over time, the above decomposition of the wealth and the hedge over the switching regimes is not well posed if we do not assume also
$\mathcal{A}^{\mathcal{R}^{def}} \bigcap \mathcal{A}^{\mathcal{R}^{fund}}  =\emptyset$. Hence being necessary in this generalized case, we have imposed it also in our case in order to give sense to the decomposition, in fact from (29)-(30) we can arrange the decomposition as follows
\[
  d\mathcal{W}_{t} =
  \begin{cases}
  d\mathcal{W}_{t}^{z} + dD^{z}_{t} - \phi^{z,mkt}_{t} d\Gamma_{t}^{\mathcal{R}^{mkt}}   = - \phi^{z,def}_{t} d\Gamma_{t}^{\mathcal{R}^{def}}  & if\; \{z_{j}=1\} \\
 d\mathcal{W}^{\zeta}_{t} +  dD^{\zeta}_{t} -\phi^{\zeta,mkt}_{t} d\Gamma_{t}^{\mathcal{R}^{mkt}}  = -  (r_{t}+ f(\mathcal{A}^{\mathcal{R}^{fund}}_{t},\mathcal{W}^{\zeta}_{t},\phi^{\zeta}_{t}))\mathcal{W}^{\zeta}_{t}dt & if\; \{z_{j}=0\}
  \end{cases}
\]

which  - by remark 2.2.4 -  we can set in particular $\phi^{z,mkt}_{t}  = \phi^{\zeta,mkt}_{t}$, so that
\[
  d\mathcal{W}_{t}  - \phi^{mkt}_{t} d\Gamma_{t}^{\mathcal{R}^{mkt}} =
  \begin{cases}
     -dD_{t}^{z} - \phi^{z,def}_{t} d\Gamma_{t}^{\mathcal{R}^{def}}  & if\; \{z_{j}=1\} \\
    -dD_{t}^{\zeta}-  (r_{t}+ f(\mathcal{A}^{\mathcal{R}^{fund}}_{t},\mathcal{W}^{\zeta}_{t},\phi^{\zeta}_{t}))\mathcal{W}^{\zeta}_{t}dt & if\; \{z_{j}=0\}
  \end{cases}
\]
which highlights the importance of [HP 1)] to give sense to our hedging representation.\\
\indent 2) As regards the assumption [HP 2)] and the self-financing condition of the hedging decomposition, they are both necessary conditions given  our objective  to price the whole contract under the martingale pricing measure $\mathbb{Q}$. In particular, the martingale condition on the \emph{discounted gain processes} $\Gamma_{t}^{i}$ ensures that also the discounted self-financing portfolio $d\mathcal{W}$ - given the initial wealth condition $\mathcal{W}_{0}$ and the hedge vector -  be a (local) martingale under $\mathbb{Q}$. \\
3) What is left to prove is that the wealth portfolio's decomposition remains self-financing. The main issue here is that the portfolio has to remain self-financing also over each switching time, which  means that at every switching time no inflow of money can  be added to the portfolio.  In fact, switching the hedging over time is not for free, given the emergence of costs by unwinding and setting up the asset positions of the relative hedging strategy. These are the so called \emph{instantaneous switching costs},  $(c_{\tau_{j}}^{z},c_{\tau_{j}}^{\zeta})$, that can be modeled as deterministic $\mathcal{F}_{\tau_{j}}-predictable$ process- for example as a fixed cost percentage of the assets values -  or stochastic ($\mathcal{F}_{\tau_{j}}-adapted$) in general\footnote{For example as a function $F(V^{\mathcal{A^{i}}}_{t},sp_{t}^{\mathcal{A^{i}}})$ of the asset value or by the market liquidity conditions. }. In the easier former case, these costs would enter the contract dividend flows, so that we would have formally
\begin{eqnarray*}
  D^{z}_{t} &:=& D^{z}_{t} + c^{z}_{t}\mathbbm{1}_{\{t=\tau_{j}\}}  \\
D^{\zeta}_{t} &:=& D^{\zeta}_{t} + c^{\zeta}_{t}\mathbbm{1}_{\{t=\tau_{j}\}}
\end{eqnarray*}
which is the underlying assumption of our wealth decomposition (29)-(30). This assumption makes thing easier because of we do not need to impose further constraints on the portfolio wealth process in order to keep it self-financing. Different and more complicate is the case of \emph{stochastic instantaneous switching costs}, which impose to check at every switching times also the validity of the self-financing condition of the portfolio hedge.\\
 A sufficiently general but strong condition that can be assumed in order to ensure the self-financing of the whole hedging strategy decomposition is the following, in addition to the initial condition on wealth:
\begin{eqnarray}
  \mathcal{W}_{0} &=& S^{C}_{0} \\
  \mathcal{W}^{z}_{\tau_{j}} &\geq& c_{\tau_{j}}^{z}, \;\; if \;\{z_{j}=1\}, \; \forall \;\tau_{j},z_{j} \in\{ \mathcal{T},\mathcal{Z}\}\\
  \mathcal{W}^{\zeta}_{\tau_{j}} &\geq& c_{\tau_{j}}^{z}, \;\; if\; \{z_{j}=0\},\; \forall \tau_{j}, z_{j} \in\{ \mathcal{T},\mathcal{Z}\}
\end{eqnarray}
where the last two conditions constrain the portfolio wealth to be almost equal to the switching costs that emerge at any optimal switching time $\tau_{j}$. $\diamond$\\

\textbf{Example 2.2.6. (Decomposition example under funding asset specification.)} Under the funding asset modeling assumptions of section 2.1, and setting as market risk and counterparty risk hedging  asset
\begin{eqnarray*}
\mathcal{A}^{\mathcal{R}^{mkt}} &=& \{\mathcal{M}_{t}^{1}, \mathcal{M}_{t}^{2}\} \\
\mathcal{A}^{\mathcal{R}^{def}} &=& \{\Lambda_{t}^{1}, \Lambda_{t}^{2},\},
\end{eqnarray*}
the hedge (30) can be formally specified as follows
\begin{eqnarray*}
 d\mathcal{W}^{z}_{t} &=& - dD^{z}_{t} + r_{t}\mathcal{W}_{t}^{z}dt + \phi^{z,\mathcal{M}^{1}}_{t} (d\mathcal{M}_{t}^{1} - r_{t}\mathcal{M}_{t}^{1}dt + d\mathcal{D}_{t}^{M^{1}})\\
 &+&  \phi^{z,\mathcal{M}^{2}}_{t} (d\mathcal{M}_{t}^{2} - r_{t}\mathcal{M}_{t}^{2}dt + d\mathcal{D}_{t}^{M^{2}})   + \phi^{z,\Lambda^{1}}_{t} ( d\Lambda_{t}^{1} - ((r_{t} + g_{t}^{\Lambda^{1}})\Lambda_{t}^{1}) dt\\
 &+& d\mathcal{D}_{t}^{\Lambda^{1}} ) + \phi^{z,\Lambda^{2}}_{t} ( d\Lambda_{t}^{2} - ((r_{t} + g_{t}^{\Lambda^{2}})\Lambda_{t}^{2}) dt  + d\mathcal{D}_{t}^{\Lambda^{2}} )
\end{eqnarray*}
\begin{eqnarray*}
 d\mathcal{W}^{\zeta}_{t} &=& - dD^{\zeta}_{t} + r_{t}\mathcal{W}^{\zeta}_{t}dt+
 ( s_{t} - bp_{t} )\mathcal{W}^{-,\zeta}_{t}dt + ( \bar{bp}_{t}-\pi_{t})\mathcal{W}^{+,\zeta}_{t}dt +  \\
   &+&   ( s_{t} - bp_{t} )\phi^{\zeta,mkt}_{t} d\Gamma_{t}^{-,\mathcal{R}^{mkt}} + ( \bar{bp}_{t}-\pi_{t})\phi^{\zeta,mkt}_{t} d\Gamma_{t}^{+,\mathcal{R}^{mkt}}\\
   &+& \phi^{\zeta,\mathcal{M}^{1}}_{t} (d\mathcal{M}_{t}^{1}- r_{t}\mathcal{M}_{t}^{1}dt + d\mathcal{D}_{t}^{M^{1}}) +  \phi^{\zeta,\mathcal{M}^{2}}_{t} (d\mathcal{M}_{t}^{2} - r_{t}\mathcal{M}_{t}^{2}dt + d\mathcal{D}_{t}^{M^{2}})
\end{eqnarray*}
for  $\;\{z_{j}=1\}$  and $ \{\tau_{j}\leq t< \tau_{j+1} \}$.\\

\textbf{Remarks 2.2.7.} Let us remark,  by proposition 2.2.5,  the price process decomposition $dS^{C}_{t}$ (as already shown in the proof) can be recast in particular as follows
\[
  d\mathcal{W}_{t} - \phi^{mkt}_{t} d\Gamma_{t}^{\mathcal{R}^{mkt}} =
  \begin{cases}
     -dD^{z}_{t} - \phi^{z,def}_{t} d\Gamma_{t}^{\mathcal{R}^{def}}  & if\; \{z_{j}=1\} \\
    -dD^{\zeta}_{t}-  (r_{t}+ f(\mathcal{A}^{\mathcal{R}^{fund}}_{t},\mathcal{W}^{\zeta}_{t},\phi^{\zeta}_{t}))\mathcal{W}^{\zeta}_{t}dt & if\; \{z_{j}=0\}
  \end{cases}
\]
which tells us that just the assets used to hedge the counterparty/default risk and the funding assets are relevant for the decomposition over the switching times, which gives us the idea for a self-financing condition weaker than (33)-(34). Defining the switching costs related to the hedging asset as follows
\begin{eqnarray*}
  c^{z,\mathcal{A}^{\mathcal{R}^{def}}}_{\tau_{j}} &=& \sum_{i=1}^{h} c^{z,\mathcal{A}^{i,\mathcal{R}^{def}}}_{\tau_{j}} \\
   c^{\zeta,\mathcal{A}^{\mathcal{R}^{fund}}}_{\tau_{j}} &=& \sum_{i=1}^{s} c^{\zeta,\mathcal{A}^{i,\mathcal{R}^{fund}}}_{\tau_{j}}
\end{eqnarray*}
 the self-financing condition becomes
\begin{eqnarray}
  \mathcal{W}_{0} &=& S^{C}_{0} \\
  \mathcal{W}^{z}_{\tau_{j}} &\geq& c^{z,\mathcal{A}^{\mathcal{R}^{def}}}_{\tau_{j}} , \;\; if \;\{z_{j}=1\}, \; \forall \;\tau_{j},z_{j} \in\{ \mathcal{T},\mathcal{Z}\}\\
  \mathcal{W}^{\zeta}_{\tau_{j}} &\geq&    c^{z,\mathcal{A}^{\mathcal{R}^{fund}}}_{\tau_{j}}, \;\; if\; \{z_{j}=0\},\; \forall \tau_{j}, z_{j} \in\{ \mathcal{T},\mathcal{Z}\}\\
\end{eqnarray}

An other important point of proposition 2.2.5 that needs further highlight is about the error/cost process $\epsilon_{t}$. By self-financing condition and by no uniqueness of the \emph{martingale pricing measure} $\mathbb{Q}$ - as already mentioned -  one needs to search  the optimal initial wealth amount $\mathcal{W}_{0}=S^{C}_{0}$ - which is our contract price - or equivalently the optimal hedging vector $\phi^{*}_{t}$ that minimizes a certain criterion, typically in the \emph{mean-variance sense}.
In other words, by the impossibility to perfectly replicate the claim value, we need to minimize the  error/cost process $\epsilon_{t}$ that enter the price process decomposition $S_{t}^{C}$, from which we can derive it explicitly as follows:
\[
  \begin{cases}
  dS^{C,z}_{t} +  dD^{z}_{t} - r_{t}\mathcal{W}_{t}dt - \phi^{z,mkt}_{t} d\Gamma_{t}^{\mathcal{R}^{mkt}} - \phi^{z,def}_{t} d\Gamma_{t}^{\mathcal{R}^{def}}  = d\epsilon^{z}_{t}  & if\; \{z_{j}=1\} \\
  dS^{C,\zeta}_{t} + dD^{\zeta}_{t} -  (r_{t}+ f(\mathcal{A}^{\mathcal{R}^{fund}}_{t},\mathcal{W}^{\zeta}_{t},\phi^{\zeta}_{t}))\mathcal{W}^{\zeta}_{t}dt - \phi^{\zeta,mkt}_{t} d\Gamma_{t}^{\mathcal{R}^{mkt}} = d\epsilon^{\zeta}_{t}   & if\; \{z_{j}=0\}
  \end{cases}
\]
which implies by price process (18-19) and wealth process decomposition (29)-(30),
\[
  d\bar{\epsilon}_{t} =
  \begin{cases}
  dS^{C,z}_{t} -   d\mathcal{W}^{z}_{t}   = dS_{t}^{rf} +dBCVA_{t}  - d\mathcal{W}^{z}_{t}  & if\; \{z_{j}=1\} \\
  dS^{C,\zeta}_{t} - d\mathcal{W}^{\zeta}_{t}  =  dS_{t}^{rf}+ dC^{fund}_{t}(t,S^{rf}_{t})   & if\; \{z_{j}=0\}.
  \end{cases}
\]
Hence our  pricing problem can be formally expressed by the following  stochastic (recursive) \emph{hedging error/costminimization problem}
\begin{equation}
\min_{\{ \{ \mathcal{T}, \mathcal{Z}\}, \bar{\phi}_{t}\}} \mathbb{E}^{\mathbb{Q}}\Big[F(\bar{\epsilon_{s}}) | \mathcal{F}_{t}\Big]
\end{equation}
where the minimization has to be run by calculating optimally also the switching time and indicators $\{ \tau_{j} \in \mathcal{T}, z_{j}\in\mathcal{Z} \}$ (in order to take in consideration the relevant dynamics and hedging strategy for the given regime) and  the functional $F(.)$ is typically represented by the mean/variance (or also both of them) of the hedging error.\\


\textbf{Remarks 2.2.8.} Let us underline an other important point in relation to the minimization problem (37). Given the decomposition of the price process and the error, one can think to run two separated minimization programs in order to derive for each regime the optimal hedges $(\phi^{*,z}_{t},\phi^{*,\zeta}_{t})$, and then subsequently run the recursive dynamic programming in order to recover the optimal switching controls $\tau_{j}^{*}, z_{j}^{*}$ taken as already known the optimal hedge vector. Formally we would have to run the following program
\begin{equation*}
1) \qquad \min_{\{  \phi_{t}^{z}\}} \mathbb{E}^{\mathbb{Q}}\Big[F(\epsilon_{s}^{z})| \mathcal{F}_{t}\Big]
\end{equation*}
\begin{equation*}
2) \qquad \min_{\{  \phi_{t}^{\zeta}\}} \mathbb{E}^{\mathbb{Q}}\Big[F(\epsilon_{s}^{\zeta})|\mathcal{F}_{t}\Big]
\end{equation*}
\begin{equation}
3) \min_{ \{ \mathcal{T}, \mathcal{Z}\}} \mathbb{E}^{\mathbb{Q}}\Big[F(\bar{\epsilon}_{s})|_{\bar{\phi}=\bar{\phi}^{*}}|\mathcal{F}_{t}\Big]
\end{equation}
This program can be run only if the optimal hedging strategy and also the asset vector dynamics do not depend   over time on the switching controls. In other words, we need to  assume that the optimal switching times strategy is \emph{separable} by the optimal  hedge vector strategy in the sense that the hedge remain optimal whatever the it is the choice of the regime. Clearly this is a strong assumption from a dynamic hedging point of view, in which at every valuation  time also the expected value of the hedge from switching and the \emph{continuation value} has to be taken in account in the  valuation. Therefore also the switching costs need which enter the dividend process, need to be small enough in order to unaffect the optimal solutions of 1) and 2). \\


\section{Contract Price-Hedge solution via System of Reflected BSDE}

\subsection{Introduction}

From the results of the past section we have highlighted the main issues related to the portfolio hedge decomposition that we have shown for our generalized  contract with contingent CSA. The aim of this section is to show the following open issue:
\begin{description}
  \item[a)] \emph{the existence of an ""arbitrage-free'' price and  of an ""optimal'' (in some specific sense) hedging strategy for our theoretical contract;}
  \item[b)] \emph{the uniqueness of the hedge decomposition of proposition 2.2.5.}
  \end{description}

 In order to tackle these tasks, we are going to show that the powerful stochastic representation approach  through  \emph{backward SDE with reflection} is the more suitable in our case. In fact, given the recursive characteristics of the price process depending on the current and future optimal switching times and costs - the so called \emph{barriers} -  one generally needs to run a backward procedure in order to evaluate over time  - via calculation of conditional expectations $\mathbb{E}^{\mathbb{Q}}[.|\mathcal{F}_{t}]$ - the convenience to switch the hedging strategy or to continue on the same regime. As we know from the former section, this task has to be run under the martingale pricing measure by minimizing a certain criterion (by no uniqueness of $\mathbb{Q}$). \\
 More formally, this means to find the triple $(S^{C}_{t}, \bar{\phi}_{t},\bar{\epsilon}_{t} )$ which represent the price process, the hedge and the error/cost process optimal solution of our problem, optimal in the sense that the variance of the error $\bar{\epsilon}_{t}$ is minimized. This triple is exactly the solution of the system of RBSDE that we are going to show in the following sections.

\subsection{Reflected BSDE problem representation}


The  BSDE approach to general contract pricing and hedging  has its roots in the seminal works of El Karoui et al. (1997) which deals with \emph{reflected BSDEs} solution and Cvitanic and Karatzas (1996) that generalize it to the \emph{doubly reflected} case and its application to \emph{Dynkin games} .  Following the work of El Karoui et al. (1997), let us recall that the reflected BSDEs are a class of BSDE whose solution $Y$ is constrained to stay above (or below) a given process, called \emph{obstacle/barrier} which is represented in our case  by the  switching condition, namely the difference between the price processes between the two regimes, including the instantaneous switching costs. The constraint action  is allowed introducing an increasing process - which is our error/cost process - which pushes the solution upwards, above the obstacle.\\
More formally, let $W = (W_{t})_{0\leq t\leq T}$ be a standard $d$-dimensional Brownian motion on a standard filtered probability space $(\Omega, \mathcal{F},\mathbb{F},\mathbb{P})$ with $\mathbb{F}= (\mathcal{F}_{t})_{0\leq t\leq T} $ the natural filtration of $W$ and given the pair of terminal condition and generator $(\xi, f)$ satisfying the following conditions\footnote{For details on BSDE theory we refer in particular to the works of Pham (2009) and Yong, Zhu (1999). }:
\begin{description}
  \item[i)] $\xi \in \mathbb{L}^{2}$ namely is square integrable and $\mathcal{F}_{T}$-measurable random variable;
  \item[ii)] $f: \Omega \times [0,T]\times \mathbb{R} \times \mathbb{R}^{d}\rightarrow \mathbb{R}$ such that $f(t,0,0)$ is $\mathbb{F}$-predictable and $f(t,y,n)$ is progressively measurable for all $y,n$ and it satisfies  a uniform Lipshitz condition in $(y,n)$, namely exists a constant $K$ such that
      \begin{equation*}
        |f(t,y_{1},n_{1})- f(t,y_{2},n_{2})|\leq K(|y_{1}-y_{2}|+|n_{1}-n_{2}|) \;\: \forall \: y_{1},y_{2}, \:\forall \: n_{1},n_{2}.
      \end{equation*}
\end{description}
In addition, let us introduce the obstacle process $(L_{t})_{0\leq t\leq T}$ which is progressively measurable and belongs to the set of continuous and square integrable processes such that
\begin{equation*}
    \mathbb{E}\Big[ \sup_{0\leq t \leq T} |L_{t}|^{2}\Big] < \infty
\end{equation*}
that is assumed $\xi\geq L_{T}$. \\
Now, we can define the solution to the reflected BSDE with terminal condition and generator $(\xi, f)$ and obstacle $L$ which is given by the triple $(Y_{t},Z_{t},K_{t})$, taking values in $\mathbb{R}$, $ \mathbb{R}^{d}$ and $ \mathbb{R}^{+} $, of progressively measurable and (predictable) square integrable processes  satisfying
\begin{eqnarray}
 \; Y_{t} &=& \xi + \int_{t}^{T}f(s,Y_{s},N_{s})+ A_{T}-A_{t} - \int_{t}^{T}N_{s}dW_{s}\;\; 0\leq t \leq T\\
 \; Y_{t} &\geq& L_{t}, \;\; 0\leq t \leq T \\
 \; 0&=&  \int_{0}^{T} (Y_{t}-L_{t})dK_{t}  .
\end{eqnarray}
with the process $A$ continuous and increasing and such that $A_{0}=0$.\\
Under this general conditions  the RBSDE (40-42) solution's existence and uniqueness is ensured and proved\footnote{For the proof of existence we refer to \emph{theorem 5.2} , for uniqueness  to \emph{theorem 4.1}  of the already mentioned work El Karoui et al. (1997).}. An other important characteristic of this stochastic tool is that the triple solution $(Y_{t},N_{t},A_{t})$ of the RBSDE admits the following \emph{Snell envelope } characterization\footnote{ The result is shown in proposition 2.3, of El Karoui et al. (1997) }
\begin{equation}
Y_{t} = ess\sup_{\nu \in\Theta} \mathbb{E}\bigg[\int_{t}^{\nu} f(s,Y_{s},N_{s})ds + L_{\nu}\mathbbm{1}_{\nu\leq T} + \xi\mathbbm{1}_{\nu= T}|\mathcal{F}_{t}\bigg]
\end{equation}
where $\Theta$ is the set of all the stopping times dominated by $T$. Hence the RBSDE solution can be connected to the solution of the optimal stopping problem (43). \\

Let us now go back to our problem. By the insights of the past section on the hedging decomposition  (proposition 2.2.5) depending on the two switching regimes, our price-hedge problem needs a generalized representation  through  \emph{system of reflected BSDE with interconnected obstacles}.\\
In order to show this, we state the following conditions (based on  proposition 2.2.5 and RBSDE definition):
\begin{description}
  \item[\textbf{A)}] let us set with $ (W^{z}_{t},W^{\zeta}_{t} )_{0\leq t\leq T}$ be two standard $l+h$-dimensional  and $l$-dimensional \emph{Brownian motion} and  with $N^{z}_{t}= \phi^{z}_{t}$, $N^{\zeta}_{t} = \phi^{\zeta}_{t}$ the $\mathcal{F}_{t}$-predictable and square integrable volatilty/hedge vector processes  taking values in $\mathbb{R}^{l+h}$ and $\mathbb{R}^{l}$;
  \item[\textbf{B)}] let us also set with   $A^{z}_{t}= \epsilon^{z}_{t}$ and $A^{z}_{t}= \epsilon^{\zeta}_{t}$ the $\mathcal{F}_{t}$-predictable, continuous and increasing processes, zero valued in $t=0$;
  \item[\textbf{C)}] we also set $L^{z}_{t}=  (Y^{\zeta}_{t} - c_{t}^{z})$ and $L^{\zeta}_{t}= (Y^{z}_{t} - c_{t}^{\zeta})$ for the two \emph{obstacle processes} which are continuous, progressively measurable and square integrable processes. In particular,\\
      - $(Y^{z}_{t}, Y^{\zeta}_{t})$ represent the \emph{price process vector} of the contract in the two regimes, which are also  progressively measurable and square integrable processes taking values in $\mathbb{R}$;\\
      - $(c^{z}_{t}, c^{\zeta}_{t})$ which are the instantaneous switching costs which are (being deterministic by hypothesis) $\mathcal{F}_{t}$-predictable and integrable;
  \item[\textbf{D)}] we are left to define the \emph{generator functions} for our system as follows:\\
        - $\Psi^{z}: \Omega \times [0,T]\times \mathbb{R} \times \mathbb{R}^{l+h}\rightarrow \mathbb{R}$, being a function of  the price process and the hedge vector process $\Psi^{z}(s,Y_{t}^{z},N^{z}_{t}) $ and for which we assume the same technical condition of point ii);\\
         - $\Psi^{\zeta}: \Omega \times [0,T]\times \mathbb{R} \times \mathbb{R}^{l}\rightarrow \mathbb{R}$, being a function of  the price process and the hedge vector process $\Psi^{z}(s,Y_{t}^{z},N^{z}_{t}) $  and for which are also valid the technical condition of point ii).
\end{description}

Assuming also for convenience the \emph{terminal condition} $\xi = 0$, let us  state the following definition.\\

\textbf{Definition 3.2.1. (Contract Price-Hedge via system of RBSDE.)} \emph{Under the technical conditions A)- D) on the processes that enter the RBSDE definition, the price-hedge for our contract,  being decomposed  as in (30)-(31) depending on the optimal switching strategy, has the following representation through system of RBSDE with interconnected obstacles:}
\[
  \begin{cases}
  Y^{z}_{t} = \int_{t}^{T} \Psi^{z}(s,Y_{s}^{z},N^{s}_{t})ds - \int_{t}^{T} N^{z}_{s}dW^{z}_{s} + A^{z}_{T} - A^{z}_{t},  \\
  Y^{z}_{t} \geq (Y^{\zeta}_{t} - c^{z}_{t}), \;\forall \: t\in[0,T]  \\
  \int_{0}^{T}( Y_{t}^{z} - (Y^{\zeta}_{t} - c^{z}_{t} ))dA^{z}_{t} = 0, \; A^{z}_{0}=0  ,& \;\forall \: t\in[0,T];
  \end{cases}
  \]

\[
  \begin{cases}
  Y^{\zeta}_{t} = \int_{t}^{T} \Psi^{\zeta}(s,Y_{s}^{\zeta},N^{\zeta}_{s})ds - \int_{t}^{T} N^{\zeta}_{s}dW^{\zeta}_{s} + A^{\zeta}_{T} - A^{\zeta}_{t},  \\
   Y^{\zeta}_{t} \geq (Y^{z}_{t} - c^{\zeta}_{t}), \;\forall \: t\in[0,T]   \\
  \int_{0}^{T}( Y_{t}^{\zeta} - (Y^{z}_{t} - c^{\zeta}_{t} ))dA^{\zeta}_{t} = 0, \; A^{\zeta}_{0}=0 ,& \; \forall \: t\in[0,T].\\
  \end{cases}
\]
\emph{and where the generator functions $\Psi(.)^{z}$, $\Psi(.)^{\zeta}$   are in particular function of the wealth decomposition (29)-(30) being function of the hedge and the price process.}\\

As already mentioned, thanks to the RBSDE representation, the solution of the RBSDE system gives us  the price, the hedge  and the hedging error/costs (two-dimensional) vectors $(\bar{Y}_{t}^{z,\zeta}, \bar{N}_{t}^{z,\zeta},\bar{A}_{t}^{z,\zeta})$,  namely namely $(\bar{S}_{t}^{z,\zeta} , \bar{\phi}_{t}^{z,\zeta},\bar{\epsilon}_{t}^{z,\zeta} )$, which are optimal in the sense of the criterion  specified by the generator function $\Psi(.)$. In particular, by assuming as criterion for the agent the \emph{hedging costs variance minimization},  the system solution will give us the contract price, the hedge and the error which are optimal under the \emph{minimizing-variance martingale pricing measure}, say $\mathbb{Q}$.\\


\textbf{Remarks 3.2.2.} In order to derive and get also the information  about the optimal sequence of switching times and indicators, we know from (43) that   the RBSDE system of definition 3.2.1 admits a solution representation via the following (non-linear) system of \emph{Snell envelope}
\begin{equation*}
    Y^{z}_{t}=ess\: sup_{\{\tau:=\tau_{j} \in\mathcal{T} \} }\mathbb{E}^{\mathbb{Q}}\bigg[\int_{t}^{\tau} \Psi^{z}(s,Y_{s}^{\zeta},N^{\zeta}_{s})ds + (Y^{\zeta}_{\tau} - c^{z}_{t})\mathbbm{1}_{\{\tau <T\}} |\mathcal{F}_{t}\bigg], \; Y^{z}_{T}=0,
\end{equation*}
\begin{equation*}
    Y^{\zeta}_{t}=ess\: sup_{\{\tau:=\tau_{j} \in\mathcal{T}  \}}\mathbb{E}^{\mathbb{Q}}\bigg[\int_{t}^{\tau} \Psi^{\zeta}(s,Y_{s}^{\zeta},N^{\zeta}_{s})ds + (Y^{z}_{\tau} - c^{\zeta}_{t})\mathbbm{1}_{\{\tau <T\}} |\mathcal{F}_{t} \bigg], \; Y^{\zeta}_{T}=0.
\end{equation*}
which is   useful in order to solve the problem as an iterative optimal stopping problem\footnote{This is shown in the literature for example in Djehiche, Hamadene (2008). } to solve under the martingale pricing measure $\mathbb{Q}$.\\

We are now in the conditions to tackle the open issue a) and b) stated above. Before resuming the main results in the following theorem on pricing and hedging of a general contract with switching  type CSA, let us state the following lemma.\\ 

\textbf{Lemma 3.2.3} \textbf{(On progressive measurable processes)} \emph{Every right-continuous and adapted process is progressively measurable.}\\

\emph{Proof}. We refer to Pascucci (2012) for the proof.\\

\textbf{Theorem 3.2.4} (\textbf{Price-hedge existence and uniqueness for switching type CSA defaultable contract}).\\
\emph{Let us assume a defaultable contract with switching type CSA and a stochastic framework as defined in section 2.1. In addition, we assume that:
\begin{enumerate}
  \item exists (but is not unique) the variance minimizing martingale measure $\mathbb{Q}$ equivalent to the real/objective one $ \mathbb{P}$;
  \item exists a primary market rich enough to set up/off  over time a self-financing strategy which admits, under HP 1 and HP 2,  a decomposition like the one derived in Proposition 2.2.5  for $d\mathcal{W}$ and $d\mathcal{S}$;
  \item  we work under the conditions A)-D) for the system of reflected BSDE set in Definition 3.2.1.
\end{enumerate}
Under these assumptions the solution of the RBSDE system defined above exists and it gives us the price-hedge and the related optimal switching times and indicators $\tau_{j}^{*}, z_{j}^{*} \in\{\mathcal{T},\mathcal{Z} \} $ for our problem, that is
\begin{eqnarray}
   \bar{N}^{*}_{t}&=& \bar{\phi}_{t}^{*} := (\phi_{t}^{*,z},  \phi_{t}^{*,\zeta}) \\
  \bar{A}^{*}_{t} &=& \bar{\epsilon}_{t}^{*} :=(\epsilon_{t}^{*,z},  \epsilon_{t}^{*,\zeta})  \\
  \bar{Y}^{*}_{t} &=&  \bar{S}_{t}^{*}:= (S_{t}^{*,z},  S_{t}^{*,\zeta})
\end{eqnarray}
where the error $\bar{\epsilon}^{*} $ is variance  minimizing,  $\bar{\phi}^{*}$ is the optimal hedging strategy and $\bar{S}^{*}$  is the price of the contract which  admits the following decomposition
\begin{equation}
    S^{*}_{t} = S^{*,z}_{t}+ S^{*,\zeta}_{t}\\
\end{equation}}

\emph{Proof.} 1) The first condition 1.  is surely  necessary (and sufficient) -  by the \emph{first fundamental theorem of the arbitrage free pricing}\footnote{We refer to  Harrison and Pliska  (1981) for details.} - in order to ensure the existence of an arbitrage-free market and hence of an arbitrage-free price for our contract. This is effectively implied by the framework assumption (also stated in condition 2.) of a primary market $\mathcal{A}^{asset}$ rich enough in order to set up an ""optimal'' replicating strategy for the price-hedge of the claim and consequently of a solution for the the RBSDE system of definition 3.2.1 . Let us mention also that by the presence of perfectly unheadgeable default/jump risk for our contract, the market is not \emph{complete} which implies that the \emph{martingale measure} is not unique (by the \emph{second fundamental theorem of the arbitrage free pricing}). Formally, under decomposition (29-30), there exist the (locally) square integrable ($\mathbb{L}_{loc}^{2}$) processes
$\lambda := (\lambda^{z} = (\lambda^{1}, \dots, \lambda^{l+h}),\lambda^{\zeta} = (\lambda^{1}, \dots, \lambda^{l})) $  such that (by \emph{Girsanov theorem})
\begin{eqnarray}
  \frac{d\mathbb{Q}}{d\mathbb{P}}|_{\mathcal{G}_{t}} &=& \frac{d\mathbb{Q}}{d\mathbb{P}}|_{\mathcal{F}_{t}} = M^{z}_{t} \\
  \frac{d\mathbb{Q}}{d\mathbb{P}}|_{\mathcal{G}_{t}} &=& \frac{d\mathbb{Q}}{d\mathbb{P}}|_{\mathcal{F}_{t}} = M^{\zeta}_{t}
\end{eqnarray}
where the first equality in both the equations derives from \emph{lemma 2.1.1} (immersion property) and where $ M^{z}_{t} $ and  $M^{\zeta}_{t}$ are \emph{exponential martingales} solving
\begin{eqnarray}
 dM^{z}_{t} &=& -M^{z}_{t}(\rho^{-1}\lambda_{t}^{z}) \cdot dW_{t}^{\mathbb{Q}}, \qquad  M^{z}_{0} =1 \\
 dM^{\zeta}_{t} &=& -M^{\zeta}_{t}(\rho^{-1}\lambda_{t}^{\zeta}) \cdot dW_{t}^{\mathbb{Q}}, \qquad   M^{z}_{0} =1
\end{eqnarray}
where $W := (W^{z}_{t},W^{\zeta}_{t} )_{0\leq t\leq T}$ are Brownian motion vector and $\rho$ the relative correlation matrix, so that
\begin{equation*}
  dW_{t}^{\mathbb{Q}} = dW_{t} - \lambda_{t}dt.
\end{equation*}
which is still a Brownian motion (with changed drift) under the equivalent martingale measure $\mathbb{Q}$.
Hence, in our problem we have these dynamics for the assets (in the two switching regimes)  under the martingale pricing measure $\mathbb{Q}$  which is (by \emph{market incompleteness}) the \emph{risk/variance minimizing} one, namely the measure  under which are determined the optimal hedge vector $\phi_{t}$ and switching controls $\{\mathcal{T}, \mathcal{Z}\}$ such that the variance of the error $\epsilon$ - namely the difference between the claim replicating portfolio $\mathcal{W}$ and the price process $S^{C}$ - is minimized.\\
2) As regards condition 2. it is surely necessary to ensure the RBSDE representation and the solution decomposition (44)-(46). Using proposition 2.2.5, it can be shown  by removing the assumptions  that the wealth decomposition (29)-(30) cannot be true and hence also the RBSDE solution decomposition (47) cannot be proved. Clearly the condition is not also sufficient, because (47) and  the RBSDE representation do not imply always the above mentioned wealth decomposition.\\
3) The third condition regards the proof that the conditions A)-D) holds for the processes entering our RBSDE representation. In particular, we need to prove that the \emph{price process vector} $(Y^{z}_{t}, Y^{\zeta}_{t})$ and the obstacle processes $L^{z}_{t}, L^{\zeta}_{t}$ are continuous, progressively measurable and square integrable processes. This is easily proved by the framework assumptions of section 2.1 on the square integrability of the processes involved in the model which are c\'adl\'ag. This condition  implies also - by lemma 2.2.3, their $\mathcal{F}_{t}$-progressive measurability, being every c\'adl\'ag process progressively measurable in particular.\\
The technical conditions on the other processes are also easily verifiable under our framework as also the \emph{Lipschitz-condition}  on the \emph{generator functions} $\Psi(.)$, given that by ii)and by choosing the variance as criterion, we have
      \begin{equation*}
        |\Psi(t,y_{1},n_{1})- \Psi(t,y_{2},n_{2})|\leq K(|y_{1}-y_{2}|+|n_{1}-n_{2}|) \; \Rightarrow,
      \end{equation*}
\begin{equation*}
        |Var(t,y_{1},n_{1})- Var(t,y_{2},n_{2})|\leq K(|y_{1}-y_{2}|+|n_{1}-n_{2}|) \; \Rightarrow,
      \end{equation*}
\begin{equation*}
         0 \leq K(|y_{1}-y_{2}|+|n_{1}-n_{2}|)
      \end{equation*}
for all the real-valued couples $y_{1},y_{2}$ and $ n_{1},n_{2}$.\\
4) Hence under these conditions the existence of the solution triple $(\bar{Y}_{t}^{*}, \bar{N}_{t}^{*}, \bar{A}_{t}^{*}) $ for the system defined in 3.2.1 is ensured as also its uniqueness, which can be proved via comparison theorem 4.   1 of El Karoui et al. (1997). In particular, this implies the uniqueness of the optimal switching strategy $\tau_{j}^{*}, z_{j}^{*} \in\{\mathcal{T},\mathcal{Z} \} $ and  consequently of the representation (47) for $S^{*}_{t}$ under the variance minimizing measure $\mathbb{Q}$.$\diamond$\\

 As corollary to this theorem and in particular by the uniqueness of the RBSDE system solution, we state the following result which solves the  issue b) highlighted at the beginning of this section.\\

\textbf{Corollary 3.2.5 (Uniqueness of hedging decomposition)}.
\emph{ The uniqueness of the solution  representation (47) for $S^{*}_{t}$ is necessary to ensure the uniqueness of the hedge decomposition (29)-(30).}\\

\emph{Proof.} The result can be easily derived by contradiction, using the uniqueness of the solution (47) which is implied by the uniqueness of the optimal switching sequences  $\tau_{j}^{*}, z_{j}^{*} \in\{\mathcal{T},\mathcal{Z} \}$ which is necessary in order to ensure a well-specified hedge decomposition, as set in proposition 2.2.5.$\diamond$\\


\textbf{Remarks 3.2.6} Let us recall that under the assumption of markovian dynamics and further technical conditions for the system, the problem admits also in this case a \emph{PDE representation } in form of \emph{variational inequalities} whose solution in viscosity sense coincide with the RBSDE one of theorem 3.2.4\footnote{Further details can be found in Crepey (2011). }. \\

\subsection{ The ""strategic interaction case'' and pricing algorithm idea}

In this ending section we are going to discuss the following  couple of important issues:
\begin{enumerate}
  \item \emph{what if we generalize our price-hedge problem allowing the strategic interaction between the party of our defaultable contract with contingent CSA of switching type;}
  \item \emph{how to practically use the theoretical results of section 3.2 on RBSDE in order to define a general pricing algorithm for a defaultable claim of our type.}
\end{enumerate}
As regards the first point, let us remind that the the analysis take over in the past section has been done under the fundamental assumption \textbf{Hp 1)} (section 2.1) that the contingent CSA is hold by just one of the party of the contract. Removing this condition and letting the contingent CSA be bilateral, the pricing/hedging problem becomes  much more complex because of the \emph{strategic interaction} tha comes in play.\\
In this generalized case, both players/counterparty can optimally manage counterparty, funding and all the contract risks  by switching from zero to full collateralization over time. This problem can be generally formulated via the theory of stochastic differential games as it has already been showed in Mottola (2013). In this work the problem is formulated in general - from the risk management and optimal design poitn of view - as a \emph{non-zero-sum stochastic differential game with switching controls} , whose existence and uniqueness of the solution, namely of a \emph{Nash equilibrium point}, remains an open problem in general.\\
From the pricing point of view, allowing the strategic interaction between the players complicates things because the price of the claim has to be the solution - in our case - of a \emph{zero-sum stochastic differential game of switching type }.
Formally we have that, setting with $J(.)$ the parties objective functionals and with $(y,u)$ the system dynamic and the relative controls set (including the switching times and idicators),  \emph{zero-sum games}  equilibrium can be defined as follows
\begin{equation}
 J^{A=B}(y, u^{*}_{A}  ,u_{B} )  \leq J^{*}(y, u^{*}_{A}  ,u^{*}_{B} )  \leq J^{A=B}(y, u_{A}  ,u^{*}_{B} )
\end{equation}
for all control sequences of $(u_{A},u_{B})$. In particular if
\begin{equation}
 \inf_{u_{A}}\sup_{u_{B}}  J^{A=B}(y, u^{}_{A}  ,u_{B} ) = \sup_{u_{B}}\inf_{u_{A}}   J^{A=B}(y, u_{A}  ,u^{}_{B} ).
\end{equation}
then the zero-sum game equilibrium -  called a \emph{saddle point } of the game - is said to have a value which represents under the \emph{martingale pricing measure} $\mathbb{Q}$, the arbitrage-free price of the claim
\begin{equation*}
    J^{*}(y, u^{*}_{A}  ,u^{*}_{B} ) := \mathbb{E}^{\mathbb{Q}}[\Pi(t,y)].
\end{equation*}
Also in this case, the solution existence and uniqueness can be studied through the theory of backward SDE given that zero-sum stochastic differential games admits a powerful stochastic representation through \emph{doubly reflected BSDE}. This implies  a generalization for the \emph{theorem 3.2.4}, which is not immediate  given that a strategic equilibrium for the game players' has to be studied,   hence we leave these issues for further research.\\
As regards the issue point b), here we want to  give just some indications about the idea of the pricing algorithm for a general defaultable claim of our type, leaving numerical applications and examples for a future paper.  Firstly, we can say that, by the multidimensional and recursive characteristic of our problem, we need
 \begin{description}
   \item[a)] a Monte Carlo procedure in order to simulate the dynamics of the number of processes involved and
   \item[b)] a backward induction procedure with the calculation of (nested) conditional expectation by projection of the solution on the available information in each step  of the recursion.
 \end{description}
 These are the main step in order to numerically implement the solution for the RBSDE defined in 3.2.1, which can be seen also (see Remarks 3.2.2) as an \emph{iterative optimal stopping procedure }by the \emph{Snell envelope } characterization  of the solution (as shown in Carmona, Ludkovski (2010)  or Mottola (2013)).\\
The point here, as from remarks 2.2.8, is to implement a numerical procedure having as output the price-hedge, namely evaluate
\begin{equation}
\min_{\{ \{ \mathcal{T}, \mathcal{Z}\}, \bar{\phi}_{t}\}} \mathbb{E}^{\mathbb{Q}}\Big[F(\bar{\epsilon_{s}}) | \mathcal{F}_{t}\Big]
\end{equation}
where the main issues involved are that the optimal hedge vector $\bar{\phi}_{t}$ has to be predictable while $\{ \mathcal{T}, \mathcal{Z}\}$ are $\mathcal{F}_{t}$-adapted and both are recursively dependent on the contract price process $S_{t}^{C}$. In addition, is worth of mention that while in market practice the hedge  is  calculated under the real/objective measure, here we assume to derive it under the martingale pricing measure.\\
Our idea of the pricing algorithm  for general defaultable claim with contingent switching CSA  can be detailed in the following main steps:
\begin{description}
  \item[1)]  discretize and simulate forward in time all the market, counterparty and funding risk dynamics and default times that enter the contract price-hedge problem, define a time-grid for  the set of controls involved and store everything;

  \item[2)] starting from the contract maturity $T$ run a backward iterative optimal stopping procedure in order to calculate over time the optimal switching times $\tau_{j}^{*}$ and regimes indicators $z_{j}^{*}$;

  \item[3)] by the knowledge of the controls of point 2) , one is able to know which is the relevant switching regime and is able to calculate the relative price process $(S_{t}^{C,z},S_{t}^{C,\zeta})$, the relative wealth $(\mathcal{W}_{t}^{z},\mathcal{W}_{t}^{\zeta})$ and hence one can calculate the optimal hedge $(\phi_{t}^{z}, \bar{\phi}_{t}^{\zeta})$ minimizing the variance of the error/cost process $(\epsilon_{t}^{z},\epsilon_{t}^{\zeta})$.

  \item[4)]  these calculations has to be performed backward over the time grid till the contract inception and finally, once it have been stored all the optimal controls, price process and errors, integrate and discount everything at the risk-free rate $r_{t}$ in order  to get in $t=0$ the initial wealth which is the contract fair value $\mathcal{W}_{0} = \mathcal{W}_{0}^{z}+\mathcal{W}_{0}^{\zeta}$ namely $S_{0}^{C} =S_{0}^{C,z} +S_{0}^{C,\zeta}$ , by the price-hedge decomposition (29-30) and (47).
\end{description}
We conclude the section by remarking - as from remarks 2.2.8 - that things become easier if  one can assume or impose the independence between the switching and the hedging control sequences. This allows to run separately the backward procedure deputed to the calculation of the switching indicators and times and the forward one which is deputed to calculate the optimal hedging vector and the variance minimizing error.


\section{Conclusions}

In this paper we have studied the problem of the  \emph{price-hedge} for general defaultable contracts characterized by the presence of a \emph{contingent CSA of switching type} whose main complication is to make the valuation recursive both in the price process and the portfolio hedge. Under fairly general framework assumptions,  and considering  in the picture the contingent bilateral CVA, collateralization (perfect vs zero) and the funding,  we have been lead to approach the solution via a \emph{portfolio hedging strategy decomposition} (proposition 2.2.5) based on the two switching regimes. Under this decompostion we have derived the solution representation for the price process of our contract through system of nonlinear \emph{reflected BSDE} whose existence and uniqueness (theorem 3.2.4) are the main contribution of the work. Based on these results, we have sketched the main steps of the pricing algorithm for our problem,  indicating also possible further research in the generalized case in which both counterparty are allowed to switch and the study of their strategic interaction  becomes central.


\newpage
\section*{\large References}





\noindent   Bielecki, T.R., Jeanblanc, M., Rutkowski, M. \emph{Hedging of defaultable claims.} Paris-Princeton Lectures on Mathematical Finance. 2004.\\

\noindent  Bouchard B. and N. Touzi. \emph{Discrete-Time Approximation and Monte-
Carlo Simulation of Backward Stochastic Differential Equations.} Stochastic
Processes and their Applications. 2004.\\




\noindent Brigo, D. A.Capponi et al. \emph{Collateral Margining in Arbitrage-Free Counterparty Valuation Adjustment including Re-Hypotecation and Netting}. ArxIv, pages 1-39, 2011.\\




\noindent Brigo, D., Pallavicini, D. Parini. \emph{Funding Valuation Adjustment:a consistent framework including CVA, DVA, collateral,netting rules and re-hypothecation.} Arxiv, 2011.\\

\noindent Carmona M., Ludkovski. M. \emph{Valuation of Energy Storage: An Optimal Switching Approach}. Quantitative finance, 2010.\\


\noindent Cesari G. et al. \emph{Modelling pricing and hedging counterparty credit exposure.} Springer Finance. Springer-Verlag, First edition. 2010.\\


\noindent Crepey S. \emph{A BSDE Approach to Counterparty Risk under Funding Constraints.} Forthcoming.\\

\noindent Cvitanic J., Karatzas I.  \emph{Backward stochastic differential equations with reflection and Dynkin games.} The Annals of probability. 1996.\\









\noindent El Karoui N., C. Kapoudjian, E. Pardoux, S. Peng and M. C. Quenez.  \emph{Reflected solutions of backward SDE's and related obstacle problems for PDE's}. The Annals of probability, 1997.\\






\noindent Gregory J.. \emph{Counterparty credit risk: The New Challenge for Global Financial Markets.} Wiley Financial Series, 2010.\\



\noindent Hamadene, S. and Zhang, J.\emph{ The Continuous Time Nonzero-sum Dynkin Game Problem and Application in Game Options}. ArXiv, pages 1-16, 2008.\\

\noindent Hamadene, S. and Zhang, J.\emph{ Switching problem and relates system of reflected backward SDEs}. Stochastic processes and their applications. 2010.\\






\noindent Harrison M., Pliska S. \emph{Martingales and stochastic integrals in the theory of continuous trading}. Stochastic Processes and their Applications. (1981).\\

\noindent  G. Mottola (2013), \emph{Switching type  valuation and design problems in general OTC contracts with CVA, collateral and funding issue}. PhD Thesis, School of Economics, Sapienza University of Rome\\

\noindent Oksendal B., A. Sulem. \emph{Applied stochastic control of jump diffusion}. Springer-Verlag, 2006.\\

\noindent Pascucci, A. \emph{PDE and martingale methods in option pricing}. Springer Verlag. 2011.\\

\noindent Pham, H \emph{Continuous-time Stochastic Control and Optimization with Financial Applications}. Springer Verlag. 2009.\\




\noindent Yong, Zhu. \emph{Stochastic controls. Hamiltonian systems and HJB equations.} Springer-Verlag. 1999 \\



\end{document}